# New polymorphic phase and rich phase diagram in the PdSe$_{2-x}$Te$_x$ system

Wenhao Liu[1], Mehrdad Rostami Osanloo[1], Xiqu Wang[3], Sheng Li[1], Nikhil Dhale[1], Hanlin Wu[1], Maarten L. Van de Put[2], William G. Vandenberghe[2], Bing Lv[1,2]*

1. Department of Physics, The University of Texas at Dallas, Richardson, Texas 75080, USA
2. Department of Materials Science and Engineering, The University of Texas at Dallas, Richardson, Texas 75080, USA
3. Department of Chemistry, University of Houston, Houston Texas 77004, USA

* to whom the correspondence should be addressed: blv@utdallas.edu

## Abstract

We report a combined experimental and theoretical study of the PdSe$_{2-x}$Te$_x$ system. With increasing Te fraction, structural evolutions, first from an orthorhombic phase (space group *Pbca*) to a monoclinic phase (space group *C2/c*) and then to a trigonal phase (space group *P-3m1*), are observed accompanied with clearly distinct electrical transport behavior. The monoclinic phase (*C2/c*) is a completely new polymorphic phase and is discovered within a narrow range of Te composition (0.30 ≤ x ≤ 0.80). This phase has a different packing sequence from all known transition metal dichalcogenides to date. Electronic calculations and detailed transport analysis of the new polymorphic PdSe$_{1.3}$Te$_{0.7}$ phase are presented. In the trigonal phase region, superconductivity with enhanced critical temperature is also observed within a narrow range of Te content (1.0 ≤ x ≤ 1.2). The rich phase diagram, new polymorphic structure as well as abnormally enhanced superconductivity could further stimulate more interest to explore new types of polymorphs and investigate their transport and electronic properties in the transition metal dichalcogenides family that are of significant interest.

## Introduction

The study of a variety of polymorphic structures of transition metal dichalcogenides (TMDs), and the discoveries of their unique properties through atomic-scale structure control, have emerged as a research frontier of science for new exciting physics as well as miniaturizing of electronic devices, energy conversion and storage[1–9]. TMD polymorphs are chemically rather simple and structured with fundamental MX$_2$ layers where M is a transition metal (such as V, Nb, Ta, Ti, Zr, Hf, Mo, W, Pd, and Pt) and X is a chalcogen atom (such as S, Se, or Te)[10]. With strong in-plane bonding and weak out-of-plane van der Waals interactions, these materials can be easily exfoliated down to atomic thickness, and thus enable bottom-up atomic-scale structure control and unique stacking/twisting to reveal new physics or novel functionalities[11–13].

The research of various structures of TMDs have a long and fruitful history. Various polymorphic structures appear starting from either the trigonal prismatic or the octahedral coordination of the metal atoms. Depending on the different stacking order of atomic planes and possible distortions, the most commonly encountered arrangements are 1T (T: trigonal), 2H (H: hexagonal), 3R (R: rhombohedral), 1T' (distorted 1T into monoclinic phase), and T$_d$ (distorted 1T into orthorhombic phase)[14–18]. The digits (1-3) indicate the number of layers in the stacking sequence in the primitive unit cell.

The different polymorphs in TMDs often display drastically different physical phenomena. For example, three different polymorphs showing distinct properties are found for the MoTe$_2$ phases: 2H-MoTe$_2$ shows semiconducting behaviors with a bandgap of about 1.1 eV, 1T'-MoTe$_2$ has a metallic behavior[19,20], and T$_d$-MoTe$_2$ is topologically nontrivial and a candidate Weyl semimetal[21,22]. Another example is found in the TaSe$_{2-x}$Te$_x$ system, where various polymorphic structures have been observed. With increasing Te concentration, the material appears as an incommensurate charge-density-wave (ICDW) 2H-TaSe$_2$ phase, as a 3R phase with a maximal superconducting transition temperature ($T_c$) of 2.4K, as a different 1T phase (also superconducting with lower $T_c$ at 0.5-0.7 K), and finally as a normal metallic monoclinic TaTe$_2$ phase[23].

Interestingly, PdS$_2$ and PdSe$_2$, less studied TMDs, have a puckered pentagonal configuration with orthorhombic space group *Pbca* shown in Fig. 1(a). The *Pbca* space group is rarely found in TMDs and hosts pentagons. Pentagons are often considered as topological defects or geometrical frustrations[24]. Pentagonal graphene and SnS$_2$ have been theoretically predicted to possess quite unique physical properties, such as unusual negative Poisson's ratio or a room-temperature quantum spin Hall insulator state[25,26]. Different from the other TMDs where the metal coordination is either trigonal prismatic or octahedral (sometimes distorted), the Pd metal configuration in PdS$_2$ and PdSe$_2$ are in fact rectangular nets. PdSe$_2$ is highly stable in air down to the monolayer limit[27]. The bandgap varies greatly from 0.5 eV in bulk to 1.37 eV in monolayers, and the thin PdSe$_2$ field-effect transistors exhibit intrinsic ambipolar characteristic and high electron mobility [28,29]. Under high pressure, bulk PdSe$_2$ transforms from a pentagonal layered structure to a pyrite type structure and superconductivity up to 13.1 K emerges, which is the highest critical temperature ($T_c$) among all the TMD materials to date[30]. On the other hand, PdTe$_2$ adopts the layered 1T structure with space group *P-3m*1 (Fig.1(b)). PdTe$_2$ has an octahedral (i.e. trigonal antiprismatic) coordination, is metallic, and is characterized by an ABC Se-Pd-Se stacking order within the layer. Dirac semimetal and superconductivity with $T_c$ of 1.65 K have been experimentally verified to coexist in PdTe$_2$[31–33].

The distinct difference in structure and electronic properties of PdSe$_2$ and PdTe$_2$ motivate us to explore the possible structural evolution/transformation, and their associated electrical transport behavior changes in the PdSe$_{2-x}$Te$_x$ system. We have demonstrated previously that the superconductivity of PdTe$_2$ can be enhanced up to 2.73 K when half of the Te in PdTe$_2$ is replaced by Se atoms[34]. Here, we study the full range of the PdSe$_{2-x}$Te$_x$ ($0 \leq x \leq 2$) system and discover a new *C2/c* polymorphic structure with similar building unit as 1T PdTe$_2$. We complement our experimental results with theoretical first principles calculations. The PdSe$_2$-PdTe$_2$ phase diagram and associated electrical transport results, including superconductivity, are presented.

## Results and discussions

### *1. Structure evolution*

Fig. 2(a) shows the X-ray diffraction patterns for PdSe$_{2-x}$Te$_x$ samples. At first glance, one can clearly observe three different types of structures with one of the corresponding characteristic peaks highlighted in the Fig. 2(a). In the Se rich region $0 \leq x \leq 0.3$, the XRD peaks shifts slightly towards lower angle, as expected with increasing Te content, and the PdSe$_{2-x}$Te$_x$ maintains the puckered pentagonal PdSe$_2$ structure with orthorhombic lattice. A second phase starts to emerge at x = 0.3, although the major phase remains the PdSe$_2$ structure, this new phase becomes more dominant when x $\geq$ 0.5 and coexists with the PdSe$_2$-type phases until x =0.6. In the composition range of $0.7 \leq x \leq 0.8$, an XRD pure quality of the new phase was obtained. With further increasing of the Te content for $0.9 \leq x \leq 2.0$, the structure transformed to the *P-3m1*(1T) structure. Fig. 2(b) and Fig. 2(c) are the Rietveld refinement results for the PdSe$_{1.9}$Te$_{0.1}$ and PdSe$_{0.6}$Te$_{1.4}$ samples. The refined lattice parameters for PdSe$_{1.9}$Te$_{0.1}$ are $a$ = 5.7687(2) Å $b$= 5.8973(6) Å $c$=7.7160(8) Å, which are slightly bigger than those of PdSe$_2$ ($a$ = 5.7410 Å $b$= 5.8660 Å $c$=7.6910 Å) as expected. The refined lattice parameters for the PdSe$_{0.6}$Te$_{1.4}$ sample are $a = b = 3.9460(9)$ Å and $c = 5.0371(5)$ Å, slightly less than that of CdI$_2$-type PdTe$_2$ as Se atoms are smaller than Te atoms.

A small crystal from the PdSe$_{1.2}$Te$_{0.8}$ sample is selected for the X-ray single crystal diffraction. The determined crystallographic parameters, refinement details, atomic coordinates, occupancies and equivalent anisotropic displacement parameters are included in Table 1. The refined ratio of Se:Te =1.23(2):0.77(2), is rather close to the nominal composition and consistent with the chemical analysis results from SEM analysis. The crystal structure of the refined PdSe$_{1.23(2)}$Te$_{0.77(2)}$ is shown in Fig. 3(a). The refined PdSe$_{1.23(2)}$Te$_{0.77(2)}$ crystallizes in a new polymorphic structure in a monoclinic cell with space group *C2/c* (#15), the chalcogen-chalcogen interlayer interactions now appear as covalent bonds, and a three-dimensional framework is visualized in Fig. 3(a). The chalcogen–Pd–chalcogen stacking is maintained similar stacking order with a fundamental building motif Pd(Se,Te)$_6$ octahedra, just like in the 1T PdTe$_2$ phase. However, the Pd(Se,Te)$_6$ octahedra are much more distorted, and severely elongated along one direction within the plane, compared to the PdTe$_6$ octahedra in PdTe$_2$, as is shown in the Fig. 4(b). As a result, the chalcogen-chalcogen distance between the adjacent stacking layers decreases down to 2.6338(5) Å in the PdSe$_{1.23(2)}$Te$_{0.77(2)}$ from the value of 3.463 Å in the 1T-PdTe$_2$ system, suggesting an effective covalent bonding interaction between the two adjacent layers. Another distinct difference of this new polymorphic phase from PdSe$_2$ and PdTe$_2$ is on the Pd metal configurations. As shown in the Fig. 3(c), the Pd metals in the new polymorphic phase PdSe$_{1.23(2)}$Te$_{0.77(2)}$ have a rhombus packing with an obtuse angle measuring ~117°, which is completely different from the rectangular Pd packing (*i.e.*, 90°) in the PdSe$_2$ and trigonal antiprismatic Pd packing (*i.e.*, 120°) in the PdTe$_2$ structure. The Pd-Pd distance is ~3.99Å, which is smaller than 4.10 Å in the PdSe$_2$ phase and 4.04Å in the PdTe$_2$ phase.

Fig. 4 shows the formation energy for PdSe$_{2-x}$Te$_x$, in the *Pbca* phase, the *C2/c* phase and the *P-3m1* phase, calculated using DFT. The theoretical calculations show that the *Pbca* phase is the most stable phase for x=0 (PdSe$_2$), for x=0.3 up to x=0.8, the *C2/c* phase appears the most stable, and for x > 0.8, the *P-3m1* phase has the lowest energy. These results are in excellent agreement with the experimentally grown material, characterized by the X-ray diffraction results shown in Fig. 2. Overall, the agreement is remarkable although a slight discrepancy between theory and experiment is observed in the x = 0.3 to x = 0.6 region where experimentally both the *Pbca* and the *C2/c* co-exist while our theoretical calculations clearly favor the *C2/c* phases. We note that the phonon contribution is a significant component to the Gibbs free energy and our theoretical results reveal that the entropy associated with the phonons is an important component in determining which phase is most stable.

As far as we know, the *C2/c* structure has not been previously reported for TMDs. This unique polymorphic structure, through this novel Pd packing and PdSe$_6$ stacking, could offer another promising platform to investigate interesting structural and electrical behaviors near the critical point of two- and three-dimensionality.

## 2. *Electronic properties*

Fig. 5 shows the changes of electrical transport properties in the PdSe$_{2-x}$Te$_x$ system upon Te doping. As one can see from Fig. 5(a), a clear crossover in electrical transport accompanied by the semiconductor-to-metal transition is observed with increasing Te content. Fig.5(b) shows the plot of ln ($\rho/\rho_{300K}$) vs (1/*T*) for a PdSe$_{1.9}$Te$_{0.1}$ sample, where thermally activated conduction behavior is observed. Two clear linear relations of ln ($\rho/\rho_{300K}$) vs. (1/*T*) are found in the range of 140-300 K and 30-69 K, indicating an energy gap of 92 meV and 13 meV, respectively. Fig. 5(c) shows the temperature dependent electrical resistivity $\rho(T)$ for the new polymorphic phase of PdSe$_{1.3}$Te$_{0.7}$. The resistivity value is weakly dependent on the temperature and shows an overall semiconducting behavior from room temperature down to 50 K. An anomalous kink arises at the temperature ~ 50K and a clear slope change of resistivity is observed afterwards. The resistivity increases more gently and has a tendency of saturation below 50 K, which may be related with the weak localizations of electrons caused by the disorders in this system.

The semiconducting behavior of the new phase is also verified with theoretical DFT calculations. Fig. 6 shows the calculated band structure and density-of-states (DOS) of PdSe$_{1.25}$Te$_{0.75}$. The DFT results show a small (15 meV) indirect bandgap for the *C2/c* phase in agreement with the experimental transport measurements. The conduction band minimum appears at the S point while the valence band maximum appears along the Γ-Y line. We also found that without the on-site electron-electron interaction (U = 4 eV) no band gap was observed.

The inset of Fig. 5(c) highlights the plot of ln *ρ* vs (1/*T*) below 50K. The curves become almost *T*-independent at low temperature between 2 and 30K, indicating the existence of a hopping process in the materials. Several general hopping models could contribute to the electrical transport in PdSe$_{2-x}$Te$_x$. One is the nearest-neighbor hopping model (NNH) where *ρ* is proportional to exp($E_A/k_BT$). Here $E_A$ is the activation energy and $k_B$ is the Boltzmann constant[35]. Another model describes variable range hopping (VRH) where *ρ* ∝ [exp[$T_0/T$]]$^{1/(1+d)}$ and often occurs between states with larger spatial distance but closer energy[36]. Here $T_0$ is a characteristic temperature and *d* is the dimensionality of the solid-state materials. VRH has been found in many disordered systems [37–39].

In our system, one can clearly observe that the slope decreases continuously with decreasing temperature, which suggests the existence of multiple conducting channels. Therefore, we utilized one NNH model together with one VRH model to fit our low temperature resistivity data below 50 K using the following equation:

$$R(T)^{-1} = R_0^{-1} + R_1^{-1} \exp[-E_1/2k_BT] + R_2^{-1} \exp\left[-(T_2/T)^{\frac{1}{3}}\right] \quad [1]$$

Where $E_1$ is the activation energy, $T_2$ is the characteristic temperature, $R_0$, $R_1$ and $R_2$ are the related resistance coefficients. Here the temperature independent $R_0^{-1}$ represent the scattering of electrons by disorders in low temperature. The extracted activation energy $E_1$ is about 2 meV. The low activation energy $E_1$ reveals inhomogeneous localization-distributions in this system and the nearly free carriers are weakly localized by the disorder. When the temperature decreases below 30 K, the VRH dominates and the characteristic temperature is extracted to be ~ 4.8×10$^4$ K. Here the characteristic temperature has a positive relation with the optimal hopping distance of the localized electrons and the yielding value is consistent with many other TMD materials[39]. We have also tried to fit out data with 1D, 2D and 3D VRH models and all of them fail to converge. When combining one NNH with a VRH model of 1D and 3D no convergence on the fit is observed either. The fact that only the 2D VRH combined with the NNH mode fits the data well, suggests that the structure of this new polymorphic phase is closer to quasi-2D rather than 3D.

Fig. 5(d) shows the temperature dependence of the Hall resistivity $\rho_{xy}$ for PdSe$_{1.3}$Te$_{0.7}$, exhibiting the new *C2/c* phase, at 10, 30, 60, 100, 150 and 200 K. Here $\rho_{xy}$ shows linearly dependent behaviors with the magnetic field. The slopes are all negative, indicating that the charge carriers are dominantly electrons near the Fermi surface. The Hall coefficient $R_H = \rho_{xy}/\mu_0 H$ estimated by the linearly fitting of $\rho_{xy}$, is shown in the inset of Fig. 4(d). $R_H$ scales almost monotonically with temperature and there are no significant changes of $R_H$ around 50 K. The charge-carrier density is estimated to be $5.12 \times 10^{19}$ cm$^{-3}$ at 10 K, which is about 2 orders smaller than the $5.5 \times 10^{21}$ cm$^{-3}$ extracted for PdTe$_2$ in Ref.[33].

Finally, Fig. 5(e) and 5(f) show the temperature dependence of the in-plane resistivity ratio $\rho/\rho_{300K}$ of PdSe$_{2-x}$Te$_x$ samples with *x* in the range $0.9 \leq x \leq 1.5$. All of these samples show metallic behavior with residual resistivity ratios (*RRR*) around 2, which are smaller than 75 of the PdTe$_2$ single crystal[40], reflecting the substantial disorder induced by Te substitution. Remarkably, in the composition range $1.0 \leq x \leq 1.2$, superconductivity suddenly occurs with an onset $T_c$ close to 2.7 K. For compositions beyond x > 1.2, no superconductivity is detected above 1.8 K (the limit of our instrument), but the samples could possibly still be superconducting below 1.8 K, as 1T-PdTe$_2$ is superconducting at 1.65 K.

## *3. Phase diagram*

Fig. 7 summarizes the rich phase diagram for the PdSe$_{2-x}$Te$_x$ system where both crystal symmetry and lattice parameters are listed. With increasing Te content x, the structure of PdSe$_{2-x}$Te$_x$ retains an orthorhombic phase up to x = 0.3. A new monoclinic polymorphic phase starts to emerge when x ≥ 0.3 and remains stable up to x = 0.8. With further doping beyond x = 0.9, the monoclinic phase becomes unstable and the samples transform to a *P-3m1* (1T) phase. Within a narrow region of *P-3m1*(1T) phase when 1.0 ≤ x ≤ 1.2, the samples show superconductivity with onset $T_c$ close to 2.7 K, as reported by us previously[34].

In conclusion, we have carried out a systematic study of isoelectronic substitution of Te for Se in PdSe$_{2-x}$Te$_x$ solid solutions. A structural evolution with increasing Te fraction from the *Pbca* PdSe$_2$ phase to the *P-3m1* PdTe$_2$ phase was observed accompanied with clearly distinct electrical transport behavior. A completely new polymorphic *C2/c* structure was discovered within a narrow range of Te composition (x = 0.3 to x = 0.8), and the *C2/c* structure has a distinct packing structure which is different from all known TMDs to date. Theoretical first principles calculations agree with the experimental findings and revealed that the phonon-contribution to the free energy is an important factor in making the *Pbca* and *C2/c* favorable phases. The *C2/c* phase showed electron-dominated charge carriers and displayed an unusual electrical conductivity behavior which could be well explained through the combined nearest-range-hopping and variable-range-hopping model. In the *P-3m1* (1T) phase region, an enhanced superconductivity emerges in a narrow range with onset $T_c$ about 2.7 K. The rich phase diagram and new polymorphic structure discovered in this system could provide a new material platform to further investigate transport and electronic properties of different types of polymorphs in the TMD structure that of significant interest.

## Methods

### A. Experimental method

The PdSe$_{2-x}$Te$_x$ crystals were synthesized using self-flux methods. Pd ingots (99.95%, Alfa Aesar), Se shots (99.999%, Alfa Aesar) and Te pieces (99.999%, Alfa Aesar) were mixed in the right stoichiometric doping ratio in an Ar glovebox with a total moisture and oxygen level less than 0.1 ppm. The source elements were loaded inside a silica tube, which was then flame sealed under vacuum and placed in a box furnace. The temperature was slowly increased to 800 °C, held for three days and followed by furnace cooling down to room temperature.

Chemical composition of the yield crystals was verified by energy-dispersive X-ray spectroscopy (EDX) on a DM07 Zeiss Supra 40 Scanning Electron Microscope. Powder X-ray diffraction (XRD) measurements were carried out at room temperature on crushed crystals using a Rigaku SmartLab X-ray diffractometer equipped with Cu-K$\alpha$ radiation. Rietveld refinement was carried out using GSAS-II[41]. Resistivity was measured using the four-probe method in a Quantum Design Physical Property Measurement System (PPMS) down to 1.8 K. Four gold wires (30 $\mu m$ in diameter) were pasted on the sample surface by silver epoxy as four probes.

The crystal structure of the new PdSe$_{1.2}$Te$_{0.8}$ phase was determined with single crystal X-ray data measured on a Bruker SMART diffractometer equipped with an Apex II area detector and an Oxford Cryosystems 700 Series temperature controller with a Mo $K\alpha$ source ($\lambda = 0.71073$Å). The collected dataset was integrated using the Bruker Apex-II program, with the intensities corrected for the Lorentz factor, polarization, air absorption, and absorption due to variation in the path length through the detector faceplate. The data was scaled, and absorption correction was applied using SADABS. The structure was solved by using the intrinsic phasing method in SHELXT and refined using SHELXL with all atoms refined anisotropically.

## B. Theoretical model

The theoretical calculations are based on Density Functional Theory (DFT) as implemented in Vienna ab initio simulation package (VASP) [42]. To correctly account for the van der Waals forces, we employed the non-local optPBE-vdW functionals proposed by Dion *et al.* [43] with parameters by Oyedele *et al.* and Klimeš *et al.*[24,44]. Electron correlation effects in the d-orbitals of Pd were accounted for in the PBE+$U$ approximation with $U = 4$ eV. To study various compositions of PdSe$_{2-x}$Te$_x$, we used 2×2×1 supercells, providing 8 compositions of $x = \left\{0, \frac{1}{4}, \frac{1}{2}, \frac{3}{4}, 1, \frac{5}{4}, \frac{3}{2}, \frac{7}{4}, 2\right\}$. The Brillouin zone was sampled with a Γ-centered Monkhorst-Pack grid of 10×10×6, 6×12×10, and 15×15×3 k-points, for the *Pbca*, *C2/c*, and *P-3m*1 structures, respectively. The electronic wave functions were expanded in a plane-wave basis with a kinetic-energy cutoff of 325 eV. The atomic positions of each composition were optimized until the force on each atom was lower than $10^{-3}$ eV/Å and the total energy was accurate to $10^{-8}$ eV.

After optimization, the intra-layer bond-length and inter-layer distance was measured. The Gibbs free energy of each structure and composition was calculated using:

$$G = E_0 - TS_{\text{vibr}},$$

where $E_0$ is the ground state energy calculated from DFT and $E_{\text{vibr}} = TS_{\text{vibr}}$ is the vibrational Gibbs free energy associated with the phonon frequencies $\omega_i$ through

$$E_{\text{vibr}} = \sum_i \left\{\frac{\hbar\omega_i}{2} + k_B T \ln\left[1 - exp\left(-\frac{\hbar\omega_i}{k_B T}\right)\right]\right\}.$$

The formation energy was calculated as the difference between the Gibbs free energy of the compound and of the elemental metals, evaluated at $T = 300$K,

$$G_{\text{form}} = G_{\text{PdSe}_{2-x}\text{Te}_x} - G_{\text{Pd}} - (2-x)\, G_{\text{Se}} - x\, G_{\text{Te}}.$$

Finally, the exfoliation energy of PdSe$_{2-x}$Te$_x$ was estimated as the difference between the ground state energies ($E_0$) of the bulk and monolayers forms per unit of surface area.

## Acknowledgments

This work at University of Texas at Dallas is supported by US Air Force Office of Scientific Research Grant No. FA9550-19-1-0037 and National Science Foundation (DMR-1921581). We also acknowledge the support from Office of Research at University of Texas at Dallas through Seed Program for Interdisciplinary Research (SPIRe) and the Core Facility Voucher Program. The project or effort depicted is sponsored by the Department of Defense, Defense Threat Reduction Agency. The content of the information does not necessarily reflect the position or the policy of the federal government, and no official endorsement should be inferred.

Table 1. Crystallographic Data, Atomic Coordinates and Equivalent Isotropic Displacement Parameters of PdSe$_{1.2}$Te$_{0.8}$

| | | | | | | |
|---|---|---|---|---|---|---|
| Temperature | | | 295 K | | | |
| Wavelength | | | 0.71073 Å | | | |
| Cell parameters | | | $a$ = 11.2450(9) Å, $b$ =4.1877(4) Å, $c$ = 6.8110(5) Å $\beta$ = 124.326(2) , $V$= 264.88(4) Å$^3$, Z=4 | | | |
| Space group | | | $C2/c$ (No. 15) | | | |
| Absorption coefficient | | | 31.834 mm$^{-1}$ | | | |
| F (000) | | | 511.4 | | | |
| $\theta$ range for data collection | | | 4.389 - 32.921 | | | |
| Reflections collected | | | 2302 | | | |
| Independent reflections | | | 487 [[R$_{int}$] = 0.0139] | | | |
| Data/restraints/parameters | | | 487/0/18 | | | |
| Goodness-of-fit on F$^2$ | | | 1.104 | | | |
| Final R indices [ I > 2$\sigma(I)$] | | | $R_1$= 0.0225, $wR_2$= 0.0623 | | | |
| R indices (all data) | | | $R_1$= 0.0228, $wR_2$= 0.0624 | | | |
| Largest diff. peak and hole | | | 3.854 and -1.249 e·Å$^{-3}$ | | | |
| atom | Wyckoff site | x | y | z | Occ. | U$_{eq}$$^a$(Å$^2$) |
| Pd1 | 4c | 3/4 | 1/4 | 1/2 | 1 | 0.0120(2) |
| Se1 | 8f | 0.60958(3) | 0.31783(9) | 0.05186(5) | 0.615(8) | 0.0133(2) |
| Te1 | 8f | 0.60958(3) | 0.31783(9) | 0.05186(5) | 0.385(8) | 0.0133(2) |

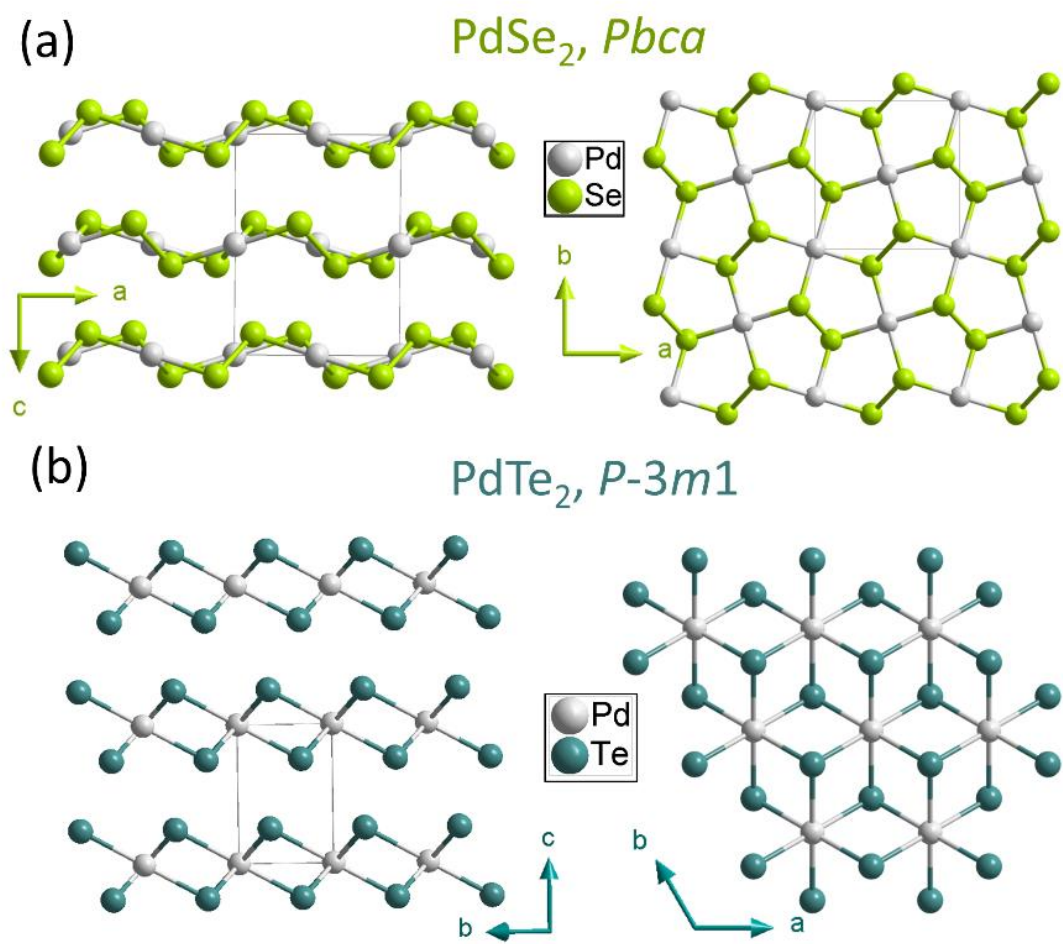

Fig. 1 Ball and stick models of (a) PdSe$_2$ and (b) PdTe$_2$ side view (left) and top view (right).

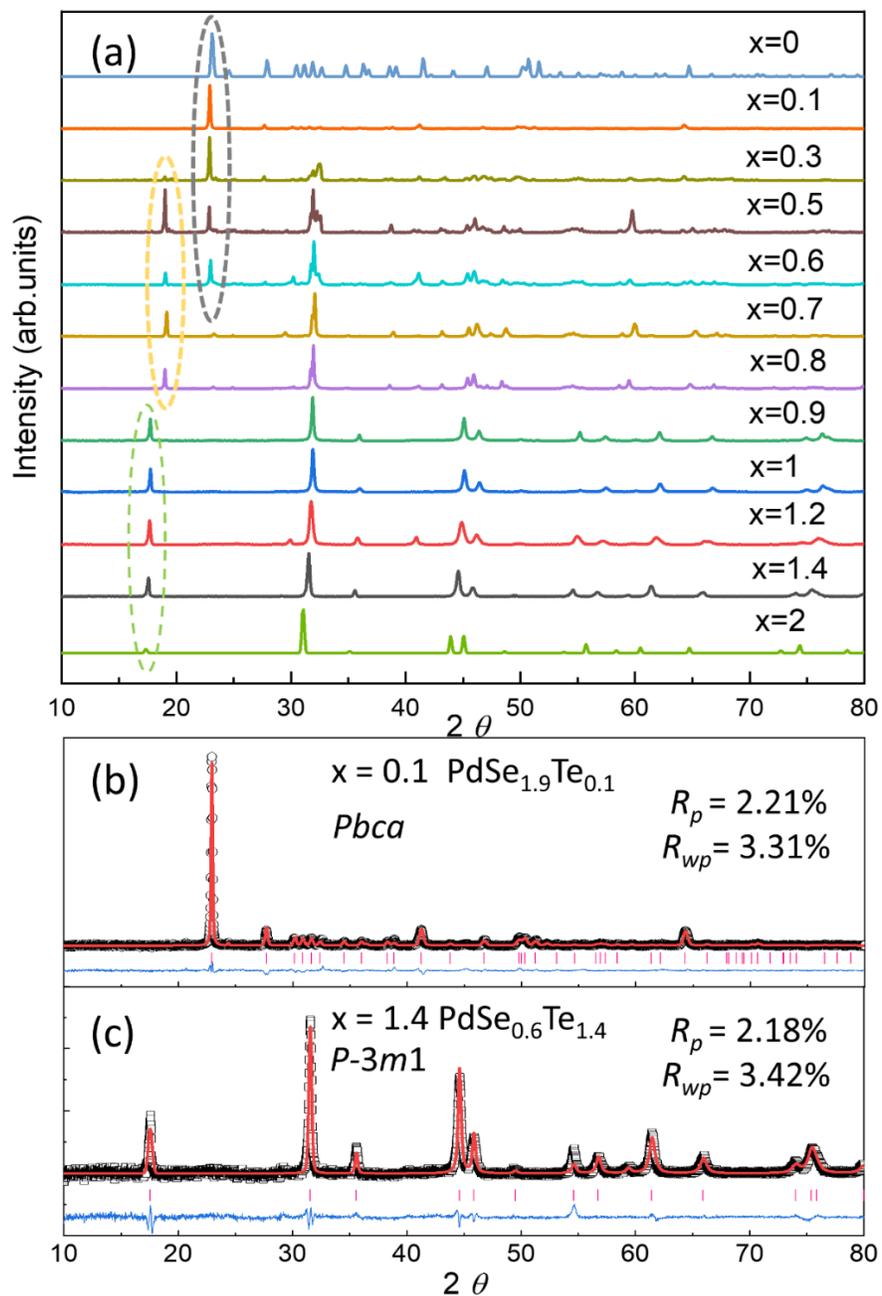

Fig. 2 (a) Powder X-ray diffraction patterns of PdSe$_{2-x}$Te$_x$ ($0 \leq x \leq 2$). (b) Rietveld refinement for PdSe$_{1.9}$Te$_{0.1}$ and (c) PdSe$_{0.6}$Te$_{1.4}$. The black circulars, red lines, pink bars, blue lines denote the observed diffraction intensities, calculated diffraction intensities, calculated locations of diffraction peaks and the difference between the calculated and observed diffraction intensities, respectively.

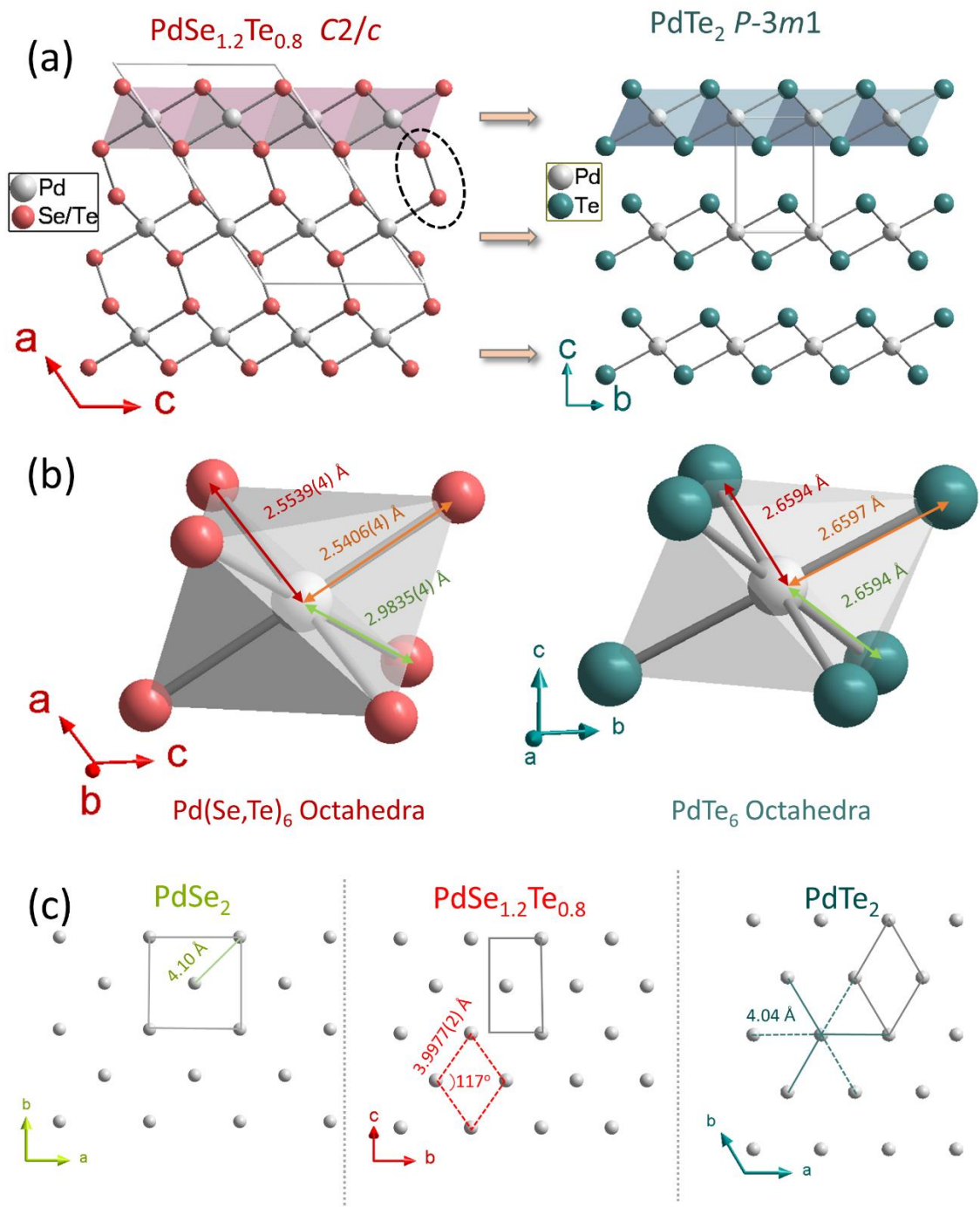

Fig. 3 (a) Ball and stick models for PdSe$_{1.2}$Te$_{0.8}$ with space group *C*2/*c* along different directions. (b) Comparison of octahedra in PdSe$_{1.2}$Te$_{0.8}$ and PdTe$_2$ structures. (c) Metal configurations in Pd planes of PdSe$_2$, PdSe$_{1.2}$Te$_{0.8}$ and PdTe$_2$.

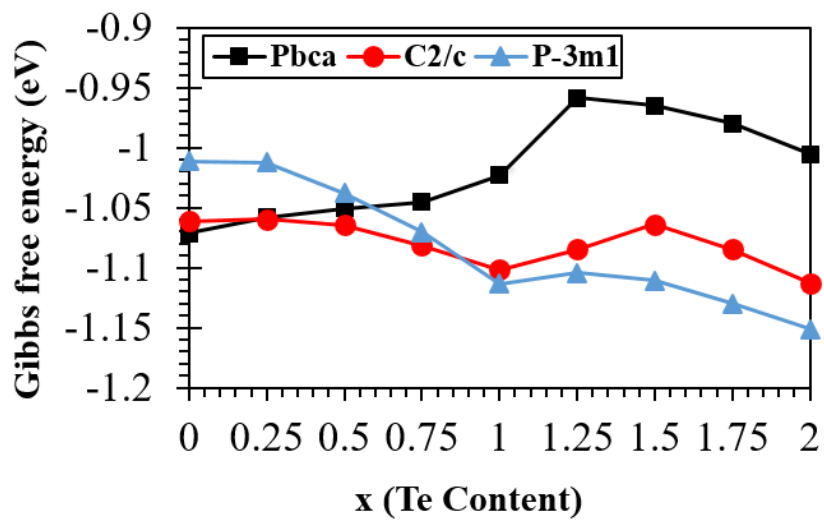

Fig. 4: Formation energy (Gibbs free energy) of *Pbca*, *C2/c* and *P-3m1* phases of PdSe$_{2-x}$Te$_x$ at 300K, calculated using DFT.

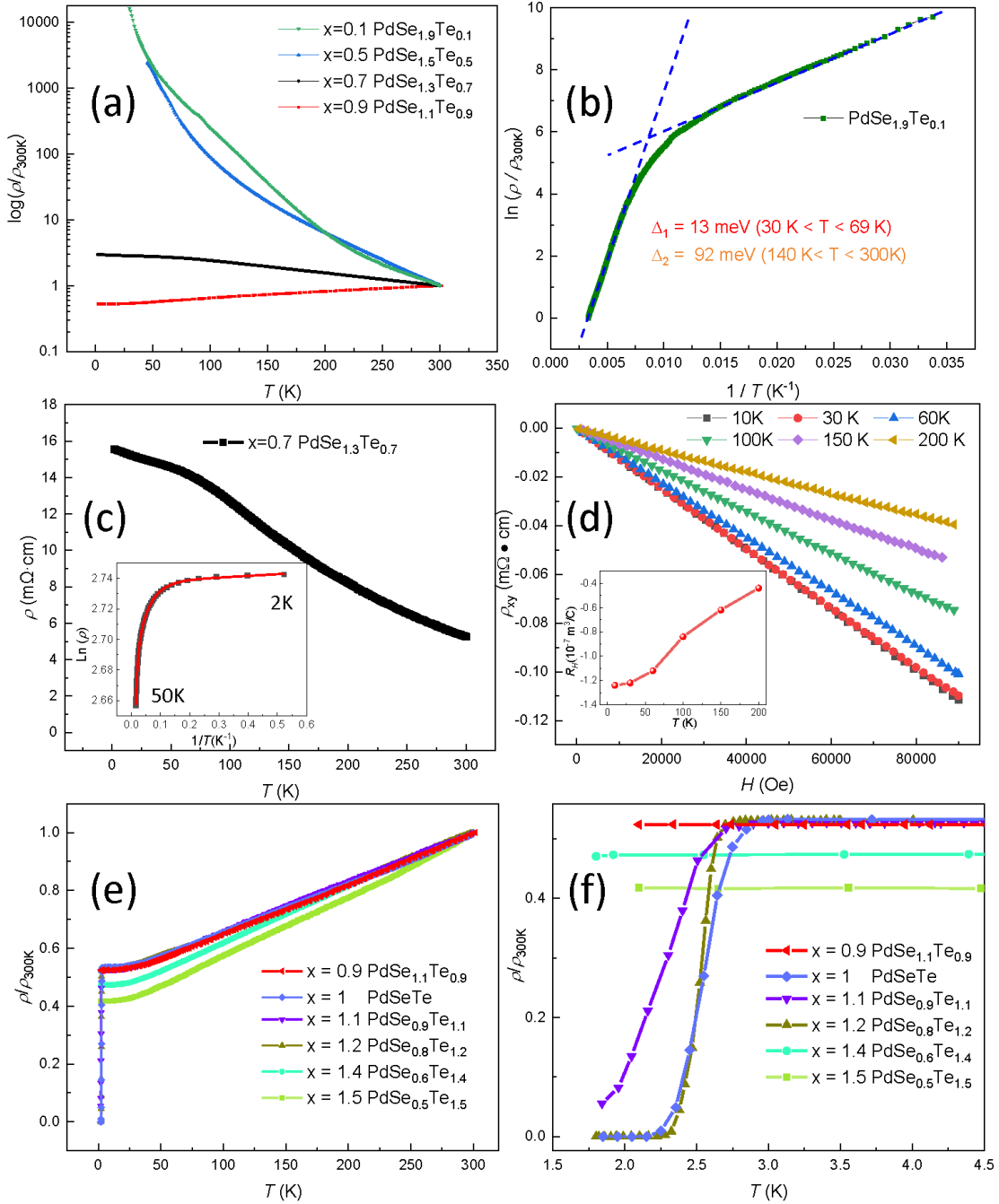

Fig. 5 (a) Temperature dependence of normalized resistivity ($\rho/\rho_{300K}$) in log scale for PdSe$_{2-x}$Te$_x$ solid solution with $0.1 \leq x \leq 0.9$. (b) Ln ($\rho/\rho_{300K}$) vs (1/$T$) for PdSe$_{1.9}$Te$_{0.1}$. The blue dashed lines represent the fitting using the standard activation model. (c) Electrical resistivity $\rho$ vs. temperature $T$ for PdSe$_{1.3}$Te$_{0.7}$. The inset shows Ln ($\rho$) vs. (1/$T$) of PdSe$_{1.3}$Te$_{0.7}$ in the temperature range 2 - 50 K. The solid red line is fitted to Eq.1. (d) Hall resistivity of PdSe$_{1.3}$Te$_{0.7}$ at 10K, 30K, 60 K, 100 K, 150 K and 200K. The inset shows the evolution of the Hall coefficient $R_H$ with temperature. (e) Temperature-dependent normalized resistance ($R/R_{300K}$) of PdSe$_{2-x}$Te$_x$ solid solution for $0.9 \leq x \leq 1.5$. (f) Enlarged view of $R/R_{300K}$ in the low temperature region 1.8 – 4.5 K.

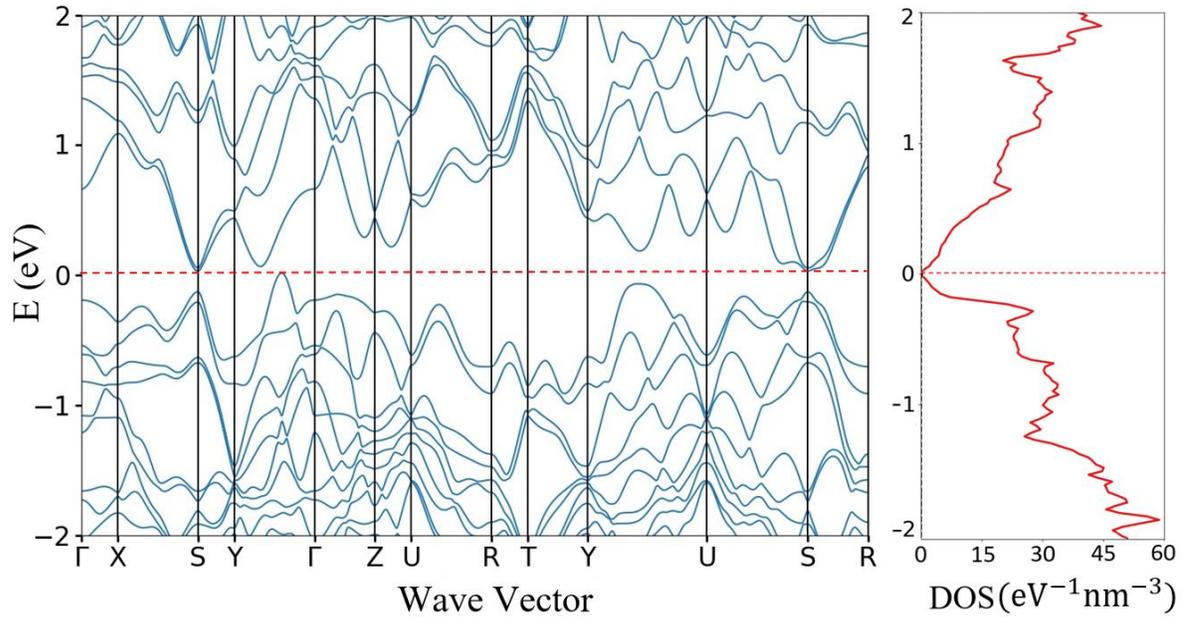

Fig. 6: Band structure and density-of-states of PdSe$_{1.25}$Te$_{0.75}$ determined from first principles.

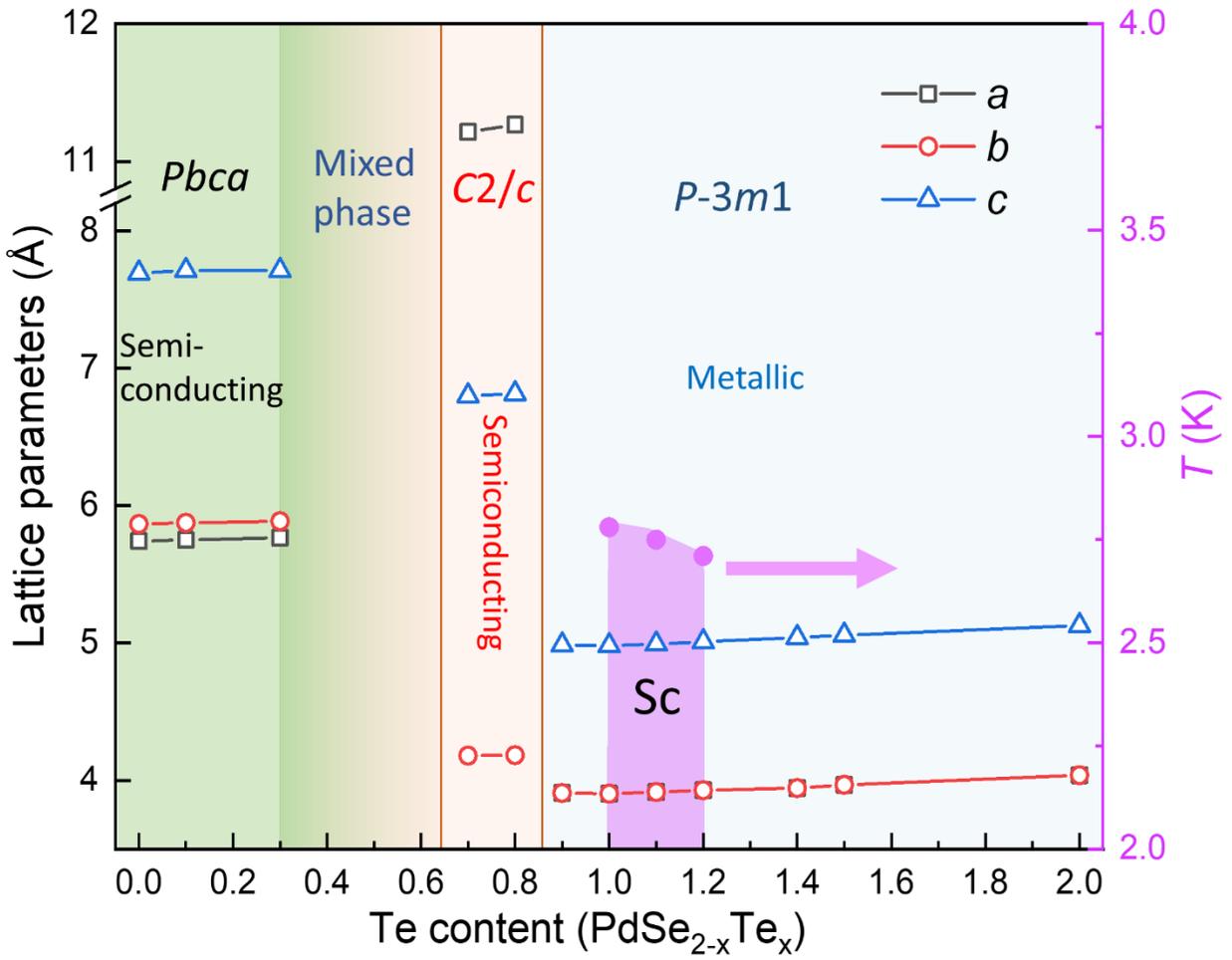

Fig. 7 Phase diagram of PdSe$_{2-x}$Te$_x$ versus Te content x. Sc denotes superconducting phase. Open squares, circulars and triangles show the lattice parameters *a*, *b* and *c* respectively. Solid purple balls show the onset superconducting transition temperature determined from transport measurements.